\documentclass{cplslarge}%
\usepackage[authoryear]{natbib}
\usepackage{rotating}
\usepackage{dcolumn}  
\usepackage{amsfonts,amssymb,amsmath} 
\usepackage{makeidx}
\makeindex
\begin{document}

\newcommand{\um}{$\mu$m}     
\newcommand{\ums}{$\mu$m }     


\newcommand{\ares}{a_{\rm res}} 
\newcommand{\nres}{n_{\rm res}} 
\newcommand{\dd}{^\circ~{\rm d}^{-1}} 
\newcommand{\dpi}{\dot\varpi}
\newcommand{\dom}{\dot\Omega}
\newcommand{\dapi}{\Delta a_{\dpi}} 
\newcommand{\daom}{\Delta a_{\dom}} 
\newcommand{\dapat}{\Delta a_{p}} 
\newcommand{\dares}{\Delta a_{\rm res}} 
\newcommand{\gcm}{{\rm g~cm}^{-2}} 
\newcommand{\cmg}{{\rm cm}^2/{\rm g}} 

\renewcommand{\thechapter}{\arabic{chapter}}    
\setcounter{chapter}{4}

\setcounter{secnumdepth}{3}

\setcounter{page}{0}
\pagestyle{empty}

~\\

\Huge
\bf
\begin{center}
The rings of Neptune  \\
\rm
\vspace{2cm}
\Large
by 
Imke~De~Pater, St\'efan~Renner, Mark~R.~Showalter, and~Bruno~Sicardy
\end{center}

\vspace{10cm}

\rm
\normalsize
This material has been published in \it Planetary Ring Systems \rm (www.cambridge.org/9781107113824) 
edited by Matt Tiscareno and Carl Murray. 
This version is free to view and download for personal use only. Not for re-distribution, re-sale or use in derivative works. 
Cambridge University Press.

\clearpage
\pagestyle{plain}

\author[de Pater et al.]{Imke~De~Pater, St\'efan~Renner, Mark~R.~Showalter, and~Bruno~Sicardy}
\chapter{The rings of Neptune}

\section{Introduction}\label{sec1}

\begin{figure}[ht]
\figurebox{10pc}{}{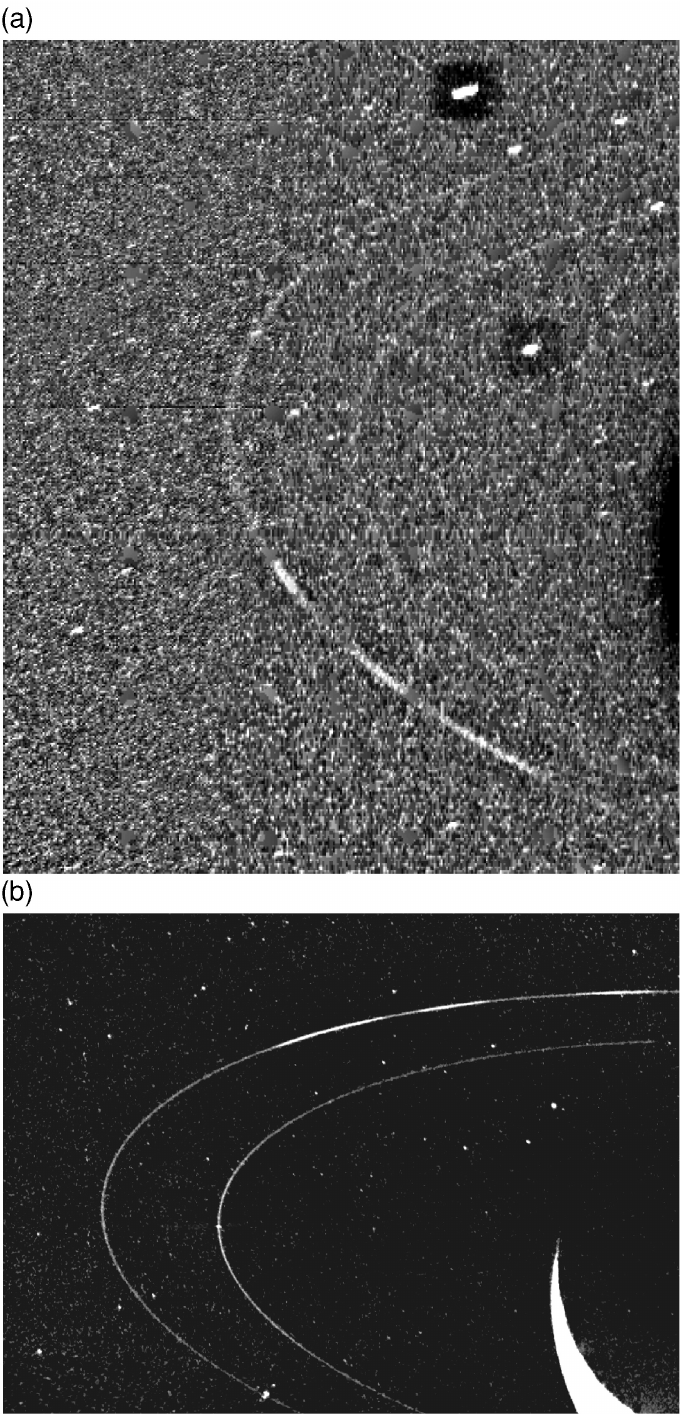}
\caption{Neptune's two most prominent rings, Adams (which includes higher
optical depth arcs) and Le Verrier, as seen by Voyager 2. (a) The rings appear faint in
this image, FDS 11350.23 (FDS numbers refer to the Voyager flight data system timeline. Each
image has a unique FDS number), taken in backscattered light (phase angle
15.5$^{\circ}$) with a resolution of 19 km per pixel. The moon Larissa
at the top of the image appears streaked as a result of its orbital motion. The other
bright object in the field is a star. The image appears very noisy because a long
exposure (111 s) and hard stretch were needed to show the faintly illuminated, low 
optical depth dark rings of Neptune. (NASA/Voyager 2, PIA00053)
(b) This forward scattered light (phase angle 134$^{\circ}$) image, FDS 11412.51, was obtained 
using a 111 second exposure with a resolution of 80 km per pixel.
The rings are much brighter in forward scattered light than in backscatter, 
indicating that a substantial fraction of the ring optical depth consists of
micrometer-size dust. (NASA/Voyager 2, PIA01493)
}
\label{fig:voyarcs}
\end{figure}

In 1984, three telescopes in South America recorded an
occultation of a star near Neptune; since the occultation was recorded
only on ingress, it was attributed to the existence of a partial ring
or ring arc \citep{Hubbard1986a, Hubbard1986b, Manfroid1986}. The
existence of ring arcs around Neptune was confirmed during subsequent
years via other occultation experiments 
\citep{Nicholson1990, Nicholson1995, Sicardy1991}, and by the Voyager~2
spacecraft \citep{Smith1989}. The Voyager observations established
that the Neptunian arcs are concentrations of particles embedded
within Neptune's narrow Adams ring, the outermost of six tenuous rings
discovered by Voyager. Figure \ref{fig:voyarcs} shows Voyager images
of the arcs in backscattered and forward scattered light.  Four ring
arcs were identified: the trailing arc Fraternit{\'e}; a
double-component arc Egalit{\'e}, dubbed Egalit{\'e} 1 and 2;
Libert{\'e}; and the leading arc Courage. The arcs varied in extent
from $\sim$~1$^\circ$ to $\sim$~10$^\circ$, and together were confined
to a longitude range of 40$^\circ$; they had typical radial widths of 
$\sim$~15 km \citep{Hubbard1986b, Sicardy1991,  
  Porco1995}. The optical depth in the arcs is of order 0.1.  Although
the arcs should have been destroyed in a few months time through
differential Keplerian motion, they appeared to persist at least
throughout the Voyager era \citep{Hubbard1986b}, and well beyond, as
discussed below.

\begin{figure}
\figurebox{13pc}{}{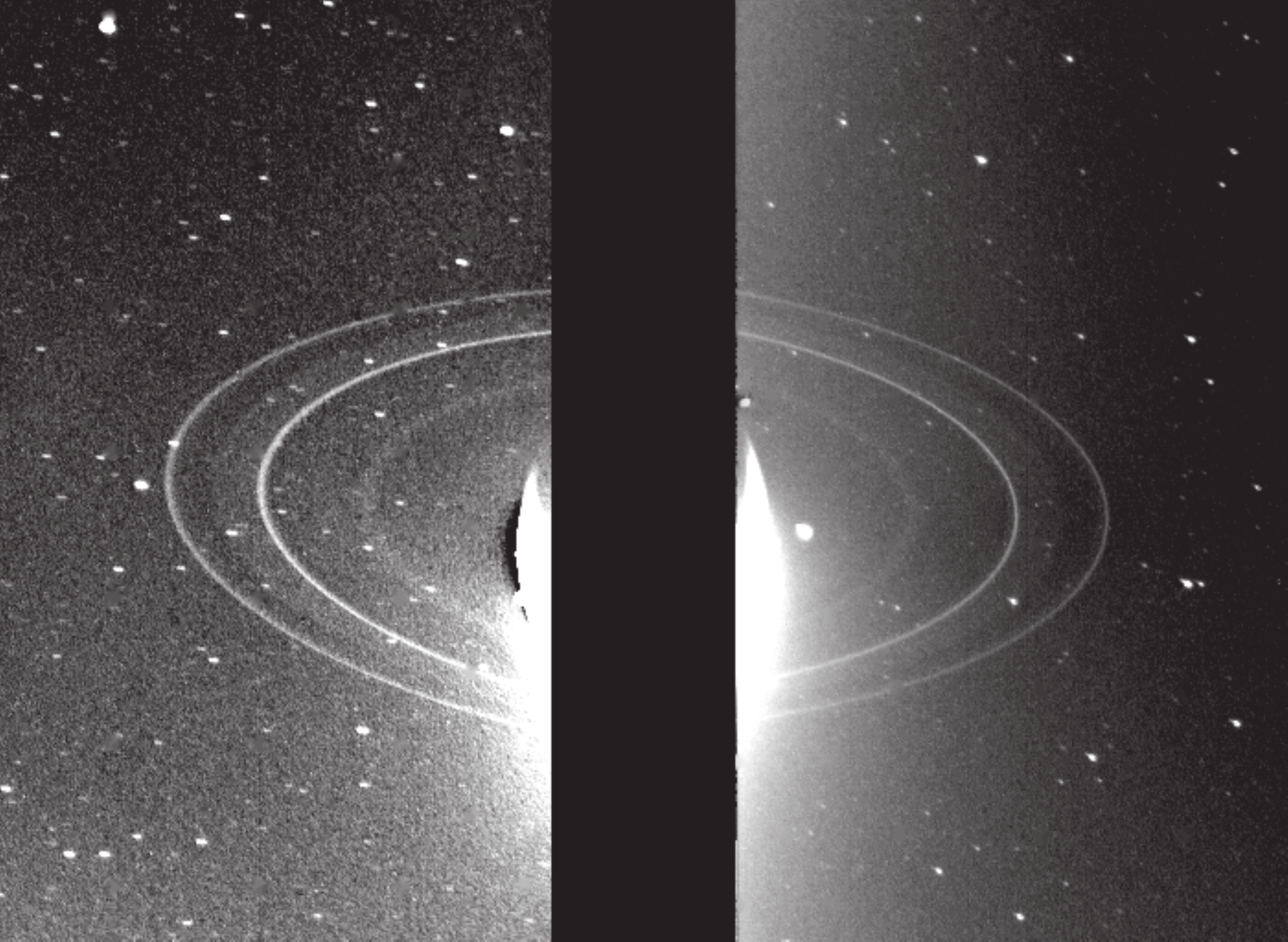}
\caption{These two 591-second exposures of the rings of Neptune were taken with the clear 
filter by the Voyager 2 wide-angle camera on Aug. 26, 1989 from a distance of 280,000 km. 
The two main rings (Le~Verrier and Adams) are clearly visible and appear complete over the region imaged. The time 
between exposures was one hour and 27 minutes. [During this period the bright ring arcs were 
on the opposite side of the planet for each exposure, such that they are not shown here.] Also visible in this image is the inner 
faint ring (Galle) at about 42,000 km from the center of Neptune, and the faint band (Lassell) that extends 
smoothly outward from the Le~Verrier ring to roughly halfway between the two bright rings. Both of these 
rings are broad and much fainter than the two narrow rings. These long exposure images were taken 
while the rings were back-lighted by the sun at a phase angle of 135$^\circ$. The bright glare in 
the center is due to over-exposure of the crescent of Neptune. (PIA01997)
}
\label{fig:voyfullrings}
\end{figure}

\begin{figure}
\figurebox{14pc}{}{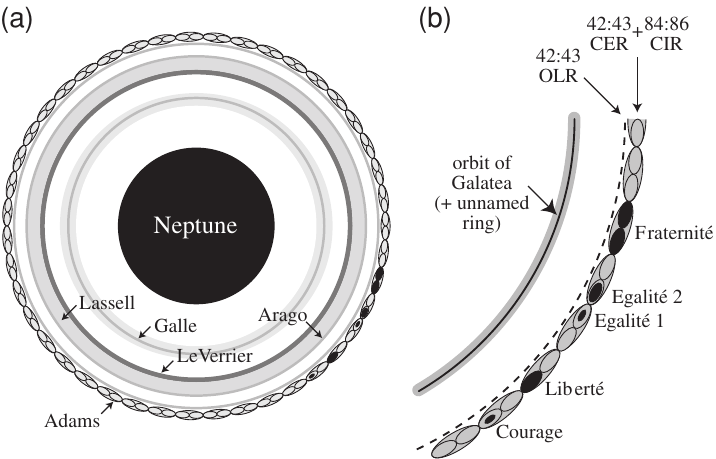}
\caption{Cartoon sketch of Neptune's rings
and associated moons as viewed from the south pole showing:
(a) the location and names of the main rings; (b) a model for the arcs in the Adams
ring (as it appeared when Voyager 2 encountered Neptune in 1989); the arcs are shown as filled centers 
of libration of the Galatea's 42:43 corotation eccentricity resonance (CER) and 84:86 corotation 
inclination resonance (CIR). The location of the two resonances (CER+CIR) as shown 
in the figure is not that of the nominal CIR but that of the observed 
arc mean motion. See Sections~\ref{sec4} and ~\ref{sec5} for details on resonances.
(Figure from \cite{dePaterLis2015}; Courtesy Carl Murray)
}
\label{fig:cartoon}
\end{figure}

The rings discovered by Voyager (Figures \ref{fig:voyfullrings},
\ref{fig:cartoon}; Table~\ref{tbl:tab1}) include the relatively bright and narrow 
Adams and Le Verrier rings, each with an optical depth of order 0.003,
the faint Lassell ring (an outward extension of the Le Verrier ring),
the Arago ring (the bright outer edge of the Lassell ring), and the
innermost faint Galle ring. Several moderately large moons orbit
within Neptune's rings (Table~\ref{tbl:tab2}); these satellites are
believed to be responsible for much of the radial and longitudinal
ring structure that has been observed. In addition to the above named
rings, there is a sixth unnamed ring, possibly incomplete, sharing the orbit of the moon Galatea. 

\begin{table*}
\caption{Properties of Neptune's ring system.$^a$}%
\begin{tabular*}{150mm}{@{}l*{6}{@{\extracolsep{0pt plus 12pt}}l}@{}}
\toprule
\vspace{-0.1cm}
&  &  &  &   & Galatea  &  \\
&Galle  & Le Verrier & Lassell & Arago  & co-orbital  & Adams \\
\hline
Radial  location  (Neptune radii) & 1.7 & 2.15 & 2.23 & 2.31 & 2.50 & 2.54 \\[4pt]
Radial  location  (km) & 42\,000 & 53\,200 & 55\,200 & 57\,200 & 61\,953 & 62\,933 \\[4pt]
Radial  width  (km) & 2000 & ${\sim}$100 & 4\,000 & & & 15 (in  arcs) \\[4pt]
Normal  optical  depth & ${\sim}$10$^{-4}$  (of  dust) & ${\sim}$0.003 & ${\sim}$10$^{-4}$& && 0.1  in  arcs \\
&&&&&&0.003   elsewhere \\[4pt]
Dust  fraction & $^b$   & ${\sim}$50$\%$ & $^b$ & $^b$ & $^b$ & ${\sim}$50$\%$  in  arcs \\[-2pt]
&  &  & & & &${\sim}$30$\%$  elsewhere \\
\botrule
\end{tabular*}
\begin{tabnote}
{$^a$ Table from \cite{dePaterLis2015}, based on data from \cite{Porco1995}.}\\
{$^b$ Large particles not detected; dust fraction is likely high.}
\end{tabnote}
\label{tbl:tab1}
\end{table*}

\begin{table*}
\caption{Neptune's inner (ring) moons$^a$}%
\begin{tabular*}{150mm}{@{}l*{6}{@{\extracolsep{0pt plus 12pt}}l}@{}}
\toprule
Satellite & a  & Orbital period & e & i  & Radius  & Albedo\\
 (\#ID)& ($10^3$ km) & (days) &  & (deg) & (km) & (geometric) \\
\hline
III Naiad & 48.227 & 0.294396 & 0.0003 & 4.69 & 33$\pm$3 & 0.07 \\
IV Thalassa & 50.074 & 0.311485 & 0.0002 & 0.14 & 41$\pm$3 & 0.09\\
V Despina & 52.526 & 0.334655 & 0.0002 & 0.07 & 75$\pm$3 & 0.09 \\
VI Galatea & 61.953 & 0.428745 & 0.0001 & 0.03 & 88$\pm$4 & 0.08 \\
VII Larissa & 73.548 & 0.554654 & 0.0014 & 0.21 & 97$\pm$3 & 0.09 \\
S/2004 N 1 & 105.3 & 0.95 & unknown & unknown & 9$\pm$1 & 0.09 (assumed) \\
VIII Proteus & 117.647 & 1.122315 & 0.0005 & 0.08 & 210$\pm$7 & 0.096 \\
\botrule
\end{tabular*}
\begin{tabnote}
{$^a$ Based on \citet{JO04} and \citet{Jake09} and http://ssd.jpl.nasa.gov; the physical parameters (radius, albedo) were based on \cite{Karkoschka2003}.}
\end{tabnote}
\label{tbl:tab2}
\end{table*}

Since the Voyager flyby, Neptune's rings have been observed with the
Hubble Space Telescope (HST), Canada-France-Hawaii Telescope (CFHT),
the Very Large Telescope (VLT), and the W. M. Keck Observatory. In the
following section we will summarize the observations and analyses
thereof that have been conducted since the Voyager flyby. In
Section~\ref{sec3} we summarize our knowledge of the particle
properties, and in Section~\ref{sec4} we review the history,
stability, and present state-of-knowledge of the arc motions. A short
review of the theories on the stability of the ring arcs is provided
in Section~\ref{sec5}. The chapter concludes with a summary,
conclusions and outlook in Section~\ref{sec6}. Chapters 12 and 13
provide background on the general physics of narrow and diffuse rings, respectively, and Chapter~13 also contains a more generalized discussion of ring arcs and confinement models.

\section{Observations since the Voyager Flyby}\label{sec2}

Images obtained with HST, the CFHT, VLT and Keck telescopes since the
Voyager flyby have consistently revealed arcs in the Adams ring. In
the following subsections we will summarize observations at
near-infrared and at visible wavelengths, and present the evolution over time of the ring arcs. To facilitate
comparison between different data sets, we typically calculate the equivalent width, $EW$,
in meters: $EW = \int (I/F) dr$, where $I$ is the observed intensity, $\pi
F$ is the solar intensity arriving at Neptune at the particular 
wavelength, and $r$ is distance across the rings.  The equivalent width is thus the equivalent extent
of a ring (in meters) if the ring had a reflectivity $I/F = 1$ (i.e., a perfect Lambertian reflector).

\subsection{Near-Infrared Observations}\label{sec2.1}

\begin{figure}
\figurebox{17pc}{}{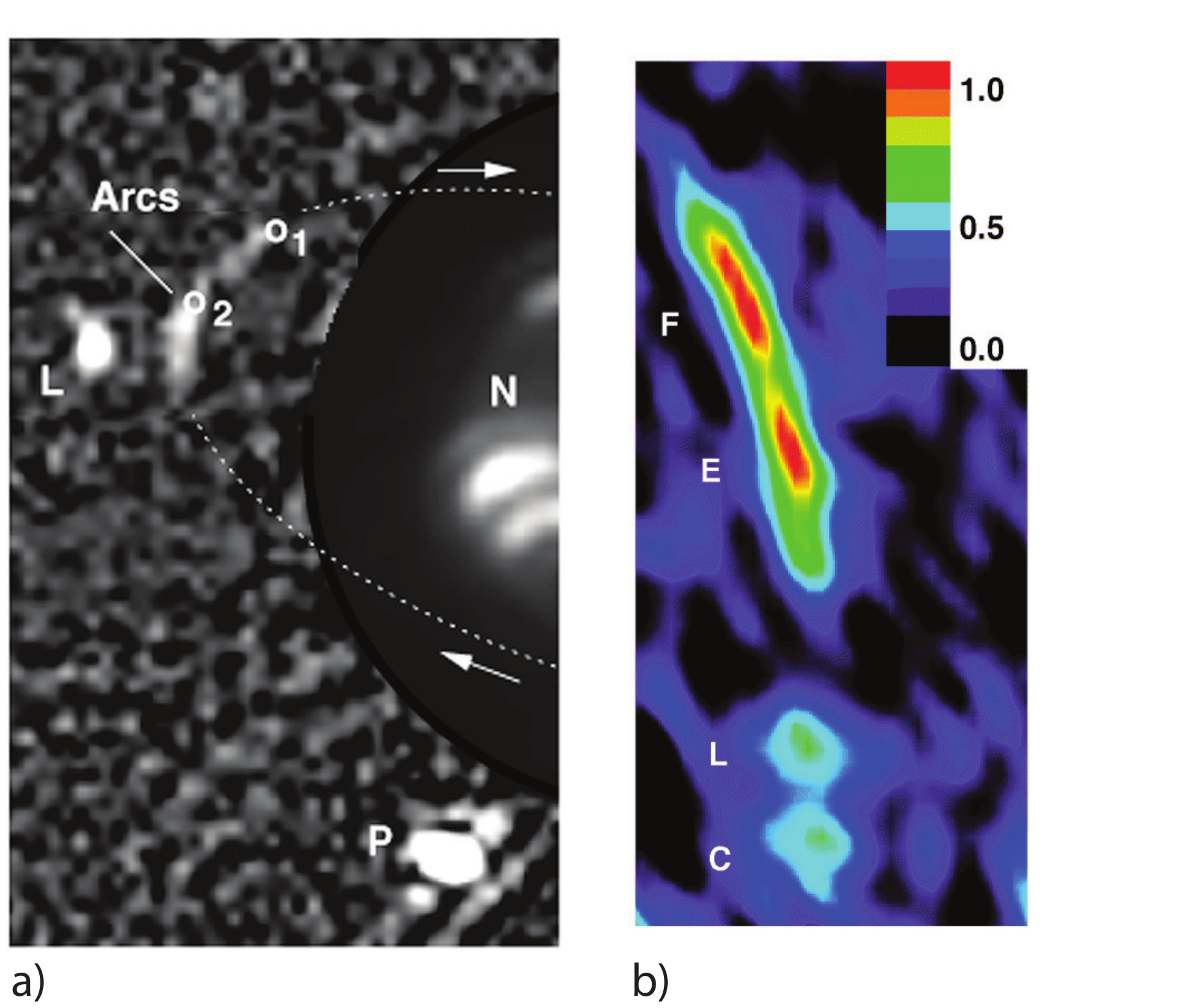}
\caption{Images of Neptune's ring arcs obtained with HST/NICMOS at 1.87 $\mu$m in 1998. 
a) a single 832-s image, where Neptune's disk was positioned partly outside the camera's field of view to
reduce its scattered light. Neptune's
contribution to the background had been modeled and removed. An image of
Neptune's clouds has been superimposed for better clarity of the system
geometry. The two circles (labeled 1 and 2) correspond to the positions of the
middle-point of the trailing arc Egalit\'e derived from two possible solutions for
the arcs' mean motion: $820.1194^\circ$/day and $820.1118^\circ$/day (Section~\ref{sec4}). The
letters L, P, and N mark the positions of Larissa, Proteus, and Neptune,  respectively,
and the arrows show the direction of motion of the ring arcs along their orbit. b)
Composite false-color image of Neptune's ring arcs, constructed from a total of 22 dithered
images, 208 s each, obtained in June and October 1998 (see text for details). The letters C, L, E and F
indicate, from leading to trailing, the respective location of the arcs Courage,
Libert\'e , Egalit\'e and Fraternit\'e. The intensity scale has been normalized to unity.
(Figure from \cite{Dumas1999}) }
\label{fig:fig_dumas}
\end{figure}

\begin{figure}
\figurebox{17pc}{}{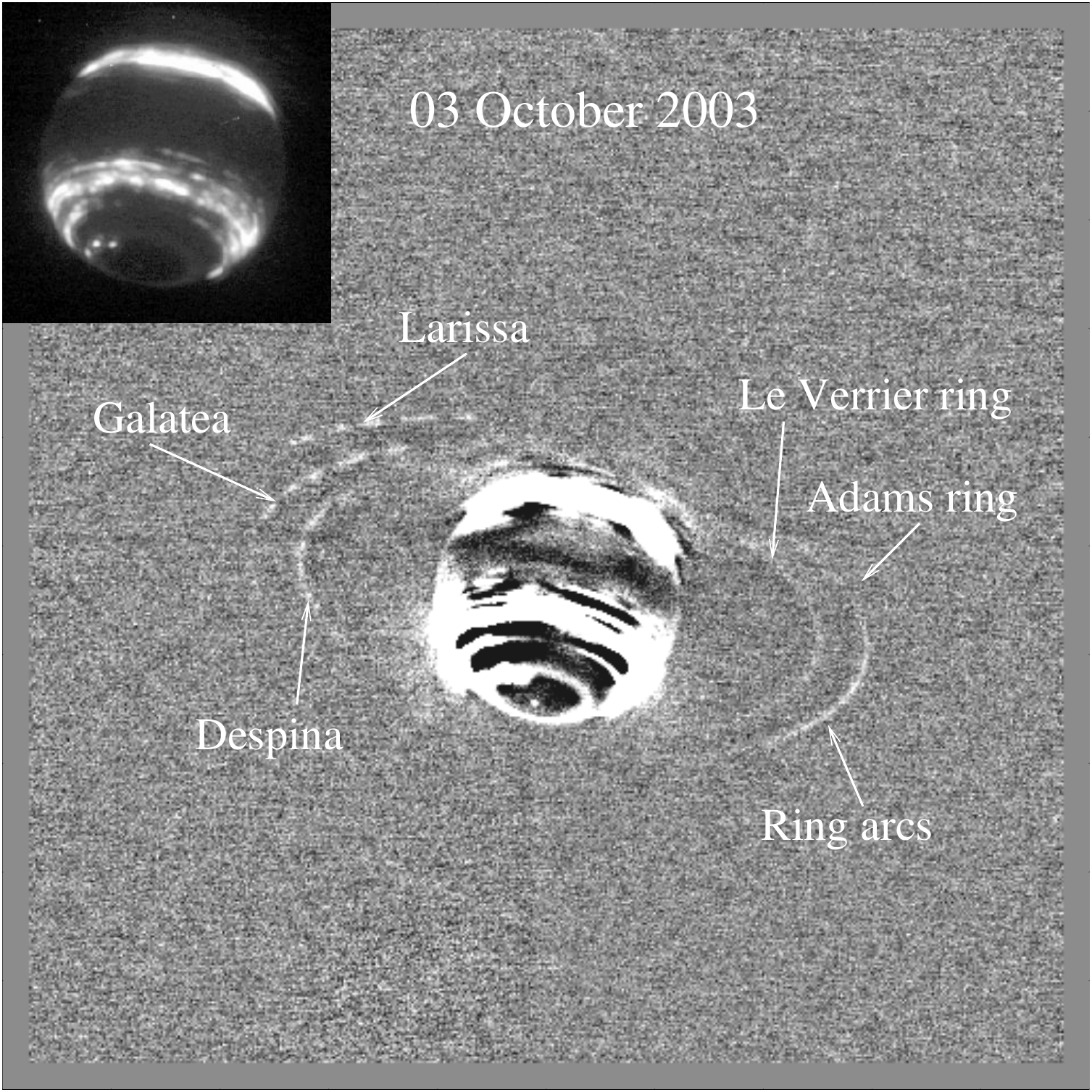}
\caption{Composite image constructed from data obtained with the Keck
  telescope at 2.12 $\mu$m on 3 October 2003, revealing satellites,
  ring arcs and the complete Adams and Le Verrier rings. The image has
  been high-pass filtered by subtracting the same image
  median-smoothed with a width of 50 pixels (1 pixel $= 0.01^{\prime\prime}$). This procedure removes
  diffuse scattered light, and brings out small-scale features
  (Neptune is highly saturated in this presentation).  A 1-minute
  exposure of Neptune itself at 2.12 $\mu$m is shown in the insert.
  (Figure from \cite{dePater2005}) }
\label{fig:fig1idp2005}
\end{figure}

\begin{figure}
\figurebox{15pc}{}{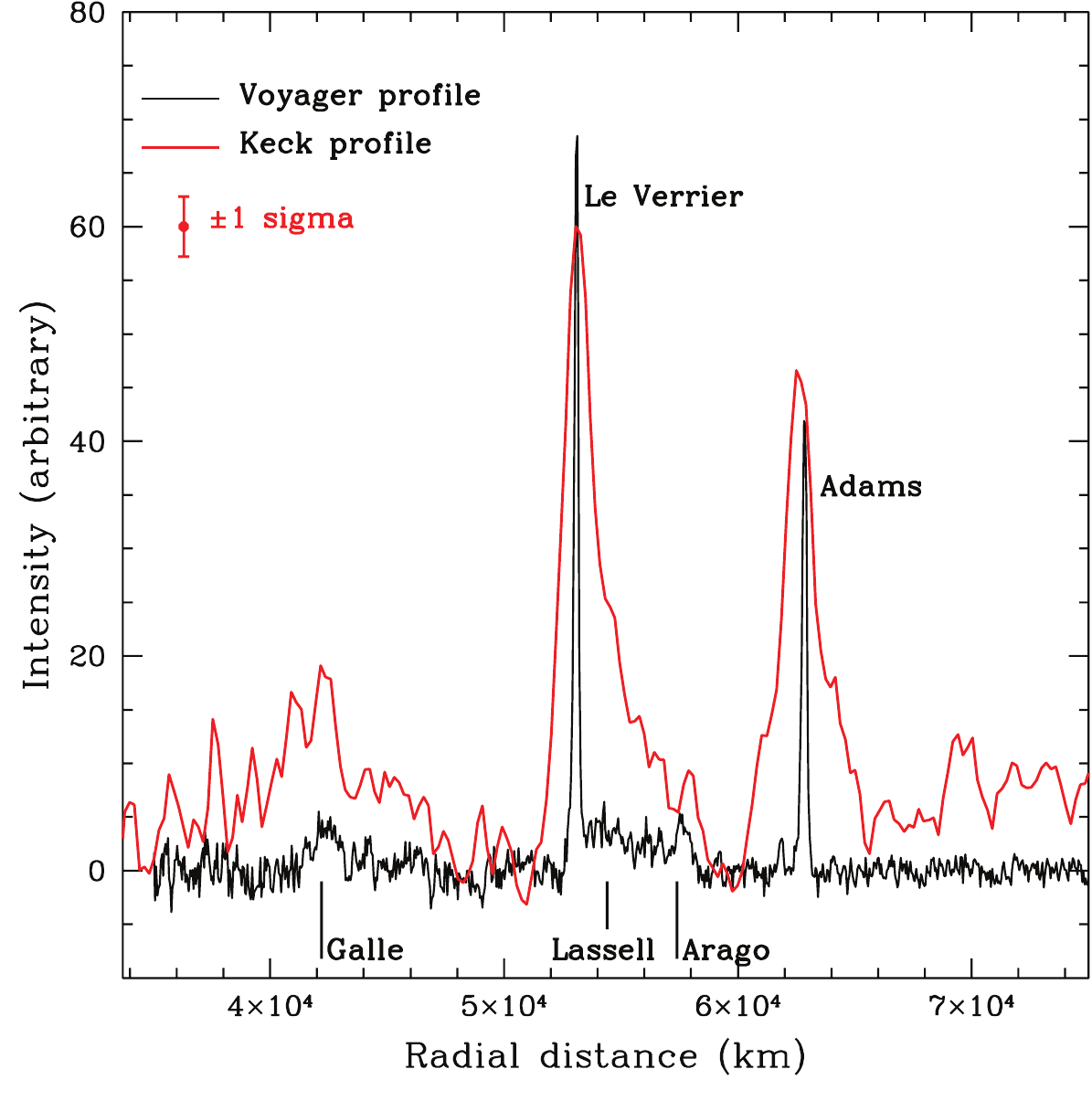}
\caption{Radial profiles through the rings. The profiles in red were
  extracted from the Keck observations in 2002 and 2003, by averaging
  over azimuth on each day (i.e., on images like that displayed in
  Fig.~\ref{fig:fig1idp2005}), away from satellites and ring arcs, and
  then averaging the profiles. The black profile is from the Voyager data.
The intensity scale is in arbitrary units, and the peak intensities of the Keck profile were scaled to those of the Voyager profile.
  (Figure from \cite{dePater2005}) }
\label{fig:fig3idp2005}
\end{figure}

\begin{figure}
\figurebox{17pc}{}{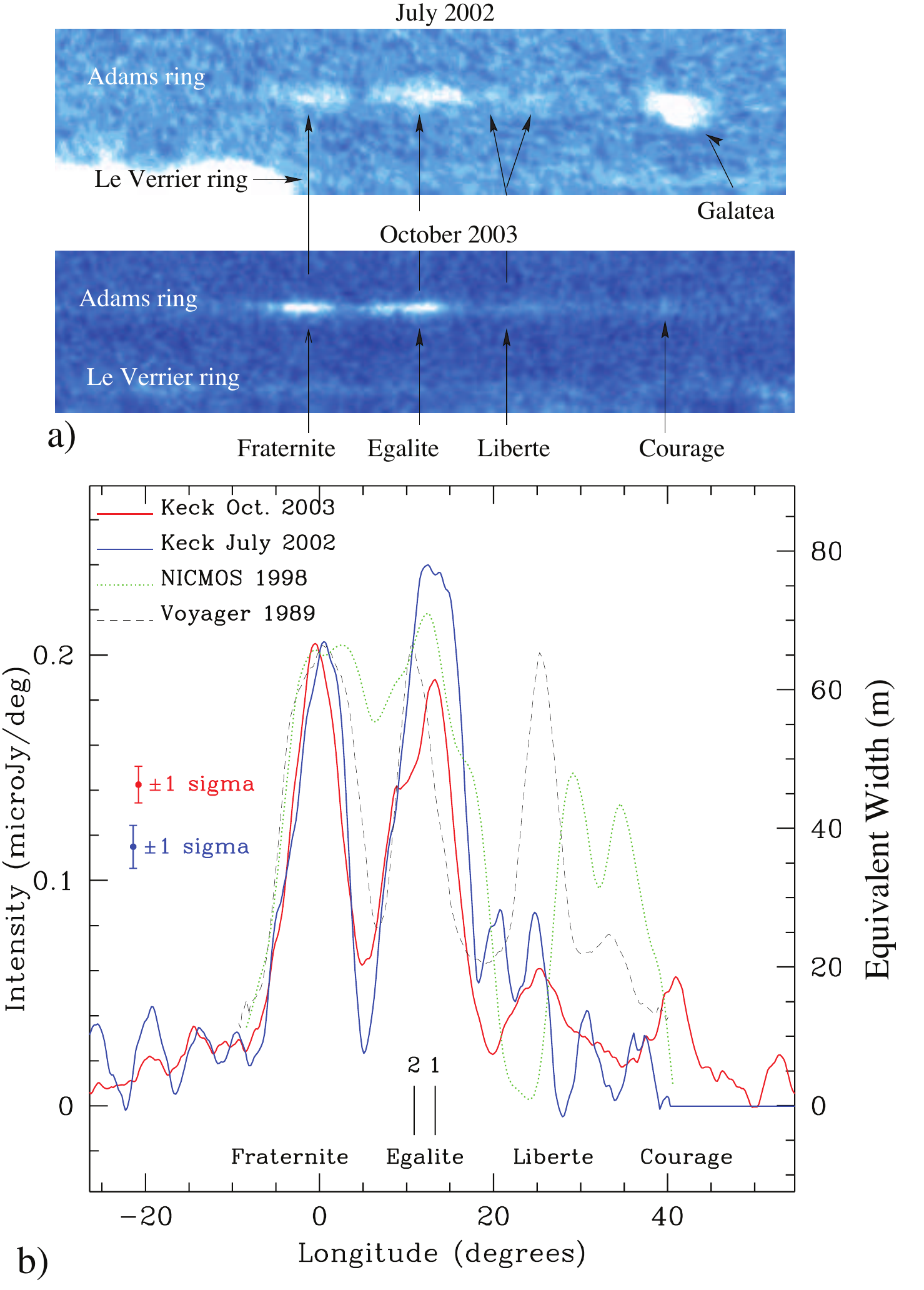}
\caption{a) Reprojected image of the ring arcs from July 2002 and October 2003. 
b) Longitudinal scans through the ring arcs as seen by Keck, Voyager (1989)
and NICMOS (1998). The Keck and Voyager profiles were smoothed to a
resolution of 3$^\circ$. All intensities are scaled to that of
Fraternit{\'e} in October 2003. Equivalent width (in m) is shown on the
right. Zero longitude was chosen to coincide with the center of arc
Fraternit\'e.
(Figure from \cite{dePater2005})
}
\label{fig:fig4idp2005}
\end{figure}

The first post-Voyager images of Neptune's ring arcs were obtained
with the Near Infrared Camera Multi-Object Spectrograph (NICMOS) on
HST \citep{Dumas1999} at 1.87 $\mu$m and from the ground with the
CFHT \citep{Sicardy1999} at 1.72
$\mu$m, both in 1998.  To minimize scattered light from Neptune, one
ideally observes the rings at wavelengths that correspond to methane
absorption bands. Light from Neptune itself is suppressed most at
1.8--2.2 $\mu$m, due to both absorption by methane gas and collision
induced absorption by hydrogen gas in the planet's atmosphere. Although
methane absorption bands are less strong at shorter wavelengths (e.g., 1.6 vs 2.1 $\mu$m), solar
radiation is stronger, and hence it is not entirely clear which
wavelength bands are best suited to detect Neptune's rings.

None of the continuous rings were
recovered in the HST/NICMOS images, but the ring arcs were. In order to
build-up sufficient signal-to-noise in any ground-based or HST near-infrared
image to identify and characterize the arcs, a large number of images
need to be combined. Due to the keplerian orbital motion of moons and
ring arcs around the planet, any co-adding of frames, or long
integration times, leads to a smearing of moons and ring arcs. 
Fig.~\ref{fig:fig_dumas}a, integrated over 832 s, is smeared over
8$^\circ$ in longitude, which corresponds roughly to the full length
of the longest arc, Fraternit\'e. Hence, in order to get a crisp image
of the ring arcs, the images need to be manipulated such as to take
out the orbital motion of the ring arcs. Fig.~\ref{fig:fig_dumas}b is
composed of 22 individual images, each integrated over 208 s. Each
image was reprojected such that the system is viewed from above, i.e.,
along the ring-plane normal, and rotated such that the arcs remained
fixed in position. The smearing effect was thus reduced by a factor of
4 compared to the image in panel a. In this composite image the ring arcs are clearly visible.

In July 2002 and October 2003 the system was imaged using the
near-infrared camera NIRC2 with the adaptive optics (AO) system on the
10-m Keck telescope at wavelengths of 1.6 $\mu$m and 2.12 $\mu$m
\citep{dePater2005}. An example of one of their composite images is
shown in Figure \ref{fig:fig1idp2005}. Each individual image was
integrated over 60 sec, so that longitudinal smearing was minimized:
i.e., in 60 sec the ring arcs are smeared by $\sim0.6^\circ$ longitude, which
corresponds to $\sim 0.035^{\prime\prime}$ (or less, depending on geometry) on the
plane of the sky, which is less than the angular resolution
of the Keck telescope ($\sim 0.05^{\prime\prime}$). The authors typically took 5
images in rapid succession. The bright short arcs in
Fig.~\ref{fig:fig1idp2005} are produced by the motion of the
satellites during this sequence of 5 images.  Both the complete Adams
and Le Verrier rings are visible in this image, which had been
integrated over a total of 30 min. Figure~\ref{fig:fig3idp2005} shows
a radial profile through the ring system (in red), compared to a
radial profile from the Voyager data (black). The Keck profile is
extracted from the 2002 and 2003 data, where on each day a radial
profile was constructed from a composite image like that displayed in
Fig.~\ref{fig:fig1idp2005}, by averaging the data in longitude away
from satellites and ring arcs. The profile shown is an average of all
profiles.  Whereas groundbased measurements of the neptunian system
are always near zero degree phase angle, the Voyager data in
Fig.~\ref{fig:fig3idp2005} were taken at a phase angle of 135$^\circ$,
highlighting dust. The brightness contrast between the Le Verrier and
Adams rings is very similar in the Keck and Voyager data, despite the
very different wavelengths and phase angles. This provides support
that both rings have similar particle properties, as had been
suggested based on the Voyager phase curves \citep{Showalter1992,
  Porco1995}.

To image the individual arcs, \cite{dePater2005} corrected for the
arcs' differential motion by ``unfolding'' the arcs along their
orbits, i.e., create images where the y-axis is along the radial
direction (i.e., from the arc to the center of planet) and the x-axis
is along the azimuthal (orbital) direction. After shifting and adding
all 51 images from July 2002, and, separately, the 93 images obtained from October~2003, high
signal-to-noise images of the ring arcs were created, which are displayed
in Figure \ref{fig:fig4idp2005}a. Longitudinal scans through these
images are shown in panel b, where the data were integrated over the
arcs in the radial direction. Although the intrinsic longitudinal
resolution is of order 0.5$^\circ$, the curves were smoothed to a longitudinal resolution of 
3$^\circ$ for better comparison 
with the Voyager and HST/NICMOS data. Equivalent width (in m) is shown on the right, scaled 
to the approximate intensity of arc Fraternit\'e in 2003.  The
approximate location of the various arcs, including the positions
of the two subregions, 1 and 2, of arc Egalit{\'e}, are indicated.  The zero-degree
longitude was chosen to coincide with the center of arc
Fraternit\'e. This figure shows that the
two leading arcs were slowly fading away, and Courage had jumped
$\sim$~8$^\circ$ ahead in its orbit by 2003, which corresponds to one full resonance site out of
43 sites (see Sections~\ref{sec4},~\ref{sec5} for explanations on resonance sites). 
The intensity of Egalit{\'e}, and
the relative intensity of its two components, varied somewhat as
well. 

\begin{figure}
\figurebox{20pc}{}{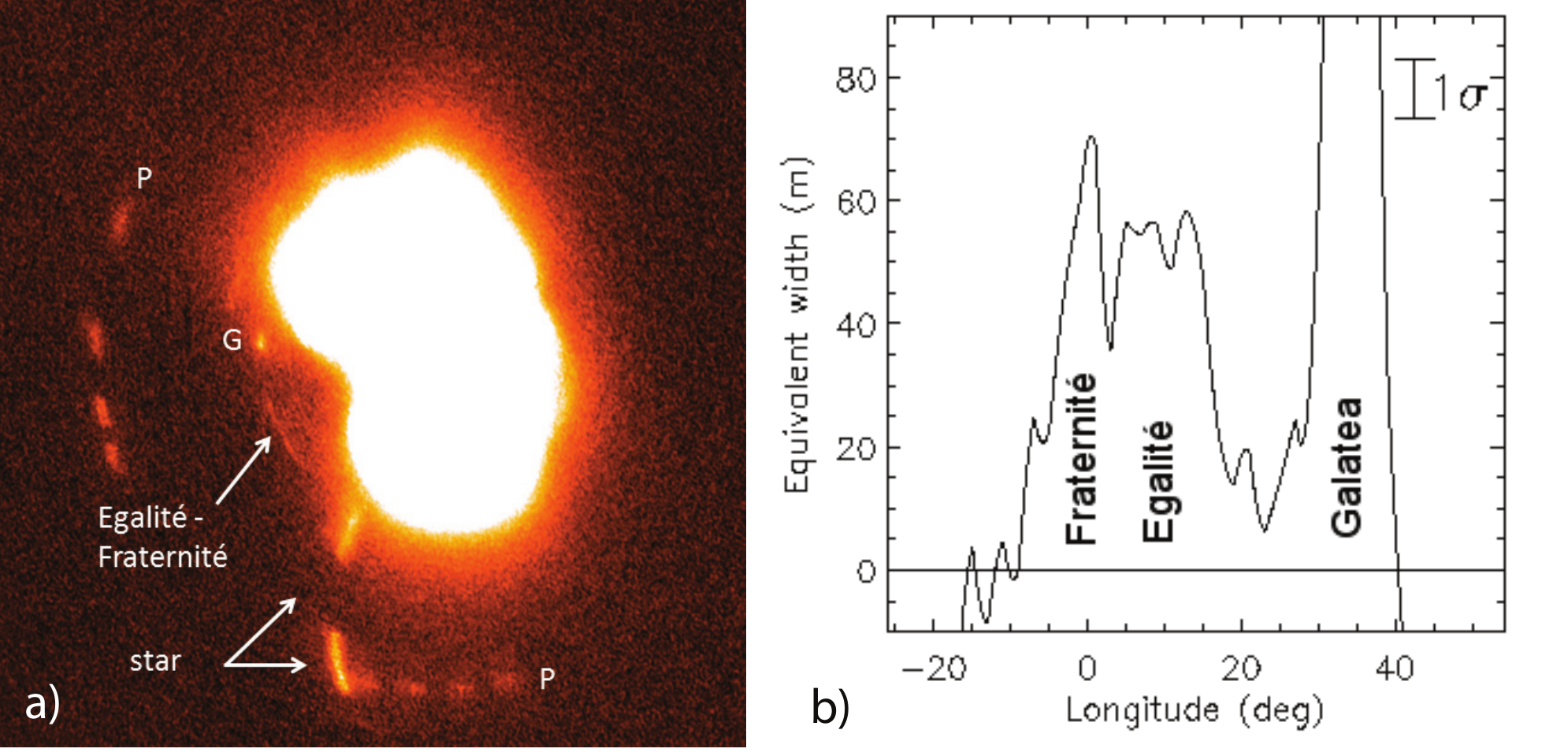}
\caption{a) Projected and co-added VLT images of Neptune's 
      equatorial plane (August 2007, 60 min total exposure time), 
      revealing material at the Fraternit\'e and Egalit\'e locations and the satellites 
      Proteus (P) and Galatea (G). Image dimension is 16.4 arcsec$^2$. b) Longitudinal 
	  scan through the ring arcs from panel a, integrated over the ring width. (Figure 
	  from \cite{Renner2014})  }
\label{VLT_coadded_image}
\end{figure}

\begin{figure}
\figurebox{20pc}{}{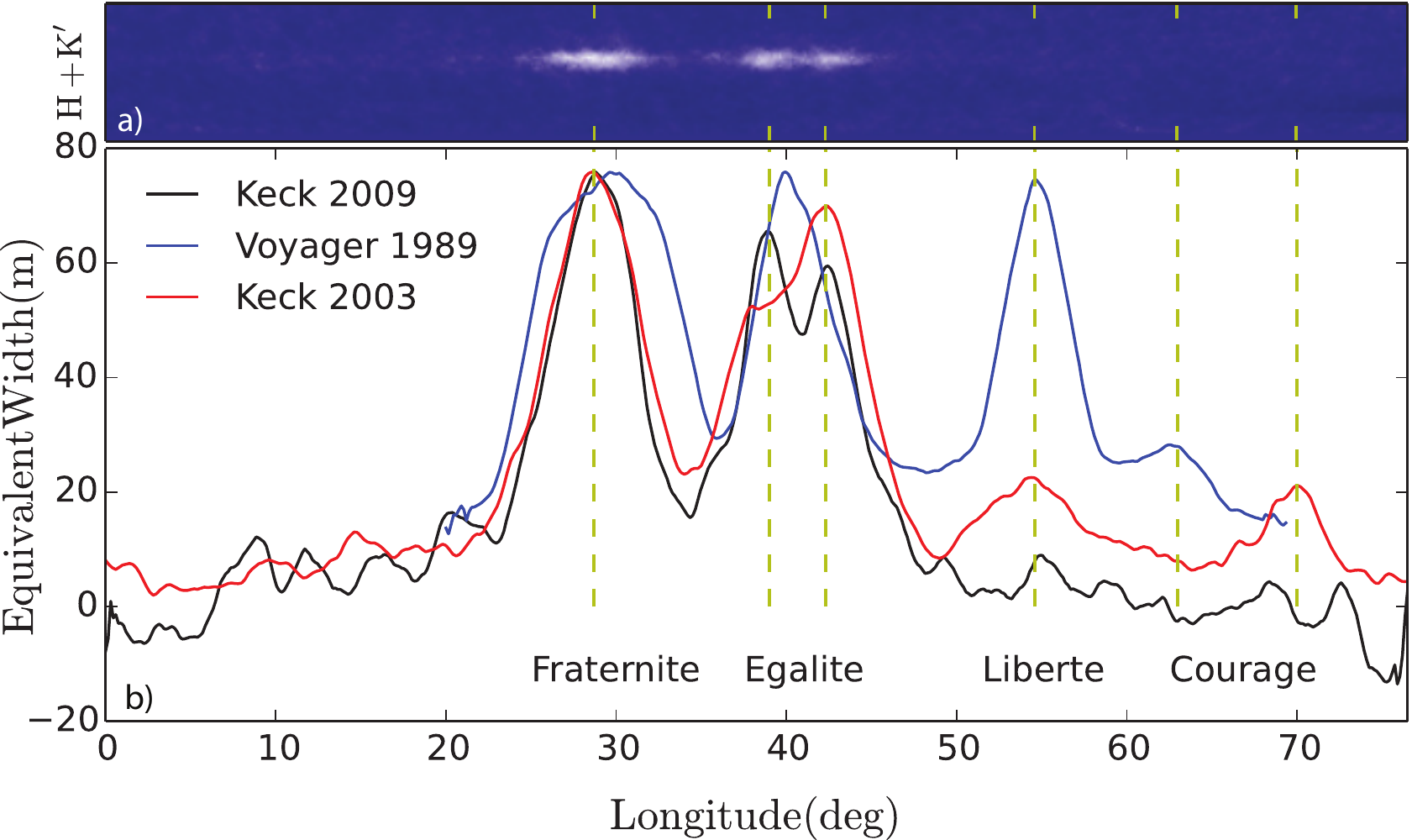}
\caption{ a) Deprojected image of the ring arcs from July 2009. Data
  at 1.63 $\mu$m (H band) were combined with data at 2.12 $\mu$m (Kp
  band). b) Longitudinal scans through the ring arcs as seen by Voyager
  in 1989, and Keck in 2003 and 2009. The Keck and Voyager profiles
  were smoothed to a resolution of 3$^\circ$. All intensities are
  scaled to that of Fraternit{\'e} in October 2003.  (Unpublished data; figure courtesy
  Patrick Fitzpatrick and Imke de Pater.) }
\label{fig:patrick}
\end{figure}

\cite{Renner2014} used the AO-fed camera NACO at the VLT to image the
Neptunian system in August 2007 at a wavelength of 2.2 $\mu$m. To
enhance the signal-to-noise of ring arcs, they reprojected the rings
in a similar way as was done with the HST/NICMOS data. Their final
image, viewed from above, is shown in Fig.~\ref{VLT_coadded_image}a. A
longitudinal scan through the ring arcs is shown in panel b.  

Figure~\ref{fig:patrick} shows the ring arcs as they appeared in July
2009, in observations obtained with the Keck telescope at 1.63 $\mu$m
and 2.12 $\mu$m; the data at these two
wavelengths were combined to produce this image and the scans. The data
reduction and analysis followed that of \cite{dePater2005}.  The data
were scaled to the arc Fraternit\'e in 2003.

While no change in Fraternit\'e's EW can be discerned between
2003 and 2007  ($\sim$~65-70 m) \citep{Renner2014}, over the years the relative
intensity of most ring arcs changed dramatically. Compared to arc
Fraternit{\'e}, Egalit{\'e} changed both in relative brightness and
longitudinal extent since the Voyager flyby. In 2002 Egalit{\'e} was
$\sim$~17\% brighter than Fraternit{\'e}, in 2003 its intensity
decreased to $\sim$~7\% below that of Fraternit{\'e}, and in 2007 and in 2009 it
was $\sim$~20\% fainter than Fraternit{\'e}.  The relative brightness
between ring arcs Egalit{\'e} 1 and 2 also changed over the years. We
further note that while Egalit{\'e} has increased in overall longitudinal
extent, Fraternit{\'e} decreased by $\sim$~25\% compared to that
observed by Voyager \citep{dePater2005}.

HST/NICMOS and Keck observations revealed that arc Libert{\'e}
decreased in intensity since the Voyager flybys.  In 1998 the arc's intensity
(relative to Fraternit{\'e}) had decreased by $\sim$~25\%, and it was observed $\sim 3^\circ$
ahead in longitude. In 2002, Libert{\'e} may have split into two narrow
arcs, separated by $\sim$~4.5$^\circ$, with the leading arc at the
original Voyager location. After smoothing the profile to the
approximate angular resolution of the NICMOS data (Fig.~\ref{fig:fig4idp2005}), the two
arcs resemble the double-hump structure observed by NICMOS; however, the NICMOS profile 
consists of the two arcs
Libert{\'e} and Courage \citep{Dumas1999}, whereas in 2002 the Keck
double-hump curve is just Libert{\'e}. In 2003, Libert{\'e} appears
again as one arc at the original Voyager (1989) location, but at just
$\sim$ 30\% of its original brightness.

Courage, usually a low-intensity arc, flared in intensity to become
nearly as bright as Libert{\'e} when NICMOS observed it in 1998; this
flaring was attributed at the time to a possible exchange of material
between the two arcs. In 2003, both Courage's and Libert{\'e}'s
intensities had decreased to levels comparable to those of Courage's
intensity in 1989.  Interesting is Courage's apparent $\sim$ 8$^\circ$
shift in the leading direction in 2003 compared to the Voyager era. If
Courage was also $8^\circ$ ahead in its orbit in 2002, it would have
been hidden by bright Galatea, which was very close to the ring arcs
at that time (Fig.~\ref{fig:fig4idp2005}a).  Unfortunately, 
Galatea also interfered with the two leading arcs at the time of the VLT
observations in 2007, and hence no unambiguous information on Courage
nor Libert\'e can be derived (Fig.~\ref{VLT_coadded_image}). By 2009, both Libert{\'e} and Courage have essentially disappeared (Fig.~\ref{fig:patrick}).

\subsection{Visible-light Observations (HST)}\label{sec2.2}

Because of the intense glare from Neptune when observing outside methane absorption
bands, only the Hubble Space Telescope (HST) is able to study Neptune's inner rings
and moons at visual wavelengths. These observations have the advantage, however, that
they cover wavelengths very similar to those from Voyager in 1989, making
direct photometric comparisons possible. This in spite of the fact that the highest
phase angle observable from Earth, $1.9^\circ$, is well below the minimum phase
angle of $\sim 8^\circ$ observed by Voyager.

Typical HST observations of the Neptune system have focused on the planet itself,
using relatively short exposures. These are useful for atmospheric studies but
are inadequate to detect the rings, arcs and small moons. Long exposures, in which
the planet itself saturates, are required. Such images of the Neptune system were
obtained by HST's Advanced Camera for Surveys in 2004--2005
\citep{Showalter2005} and by the Wide Field Camera 3 (WFC3) in 2009
\citep{Showalter2013}. The analysis of a similar 2016 data set is not yet complete
\citep{Showalter2016}.

The HST observations have been notable for a few reasons that may or may not be related to the rings.
For one, these data revealed a small moon, S/2004 N 1, orbiting between Larissa
and Proteus \citep{showaltercbet3586}. Indications so far are that this moon was too small to
have been detected by Voyager's cameras. Furthermore, they have revealed that
Naiad, the innermost known moon, has deviated substantially from its predicted
orbit. Whether this reveals the influence of some hitherto unknown perturbation,
or is just the result of ephemeris uncertainty, is not yet known.

The arcs show up clearly in these images, as do the Le Verrier and Adams rings.
As was also noted in the infrared studies (Fig.~\ref{fig:patrick}), the leading arcs have dissipated and
can no longer be seen. The two trailing arcs, Egalit{\'e} and Fraternit{\'e}, remain,
although they appear to have narrowed slightly. Careful photometry suggests that
they are diminished in brightness relative to the Voyager flyby in 1989.

\section{Particle Properties}\label{sec3}

\begin{table*}
\caption{Reflectivities of Galatea and the Ring Arcs }%
\begin{tabular*}{150mm}{@{}l*{6}{@{\extracolsep{0pt plus 12pt}}l}@{}}
\toprule
  &Voyager$^a$ &NICMOS$^b$ & Keck 2002$^c$  & Keck 2003$^c$ & Keck 2003$^c$ \\
  &0.5 $\mu$m& 1.87 $\mu$m& 2.12 $\mu$m& 2.12 $\mu$m& 1.63 $\mu$m \\
\hline
Galatea I/F$^d$ & 0.079 &0.086 $\pm$ 0.06  &0.096 $\pm$ 0.015  & 0.094 $\pm$ 0.011 & 0.088 $\pm$ 0.012 \\
Arcs (F$+$E) I/F$^d$ &0.055 $\pm$ 0.004 & 0.083 $\pm$ 0.012& & 0.088 $\pm$ 0.012 & 0.082 $\pm$ 0.013 \\
\botrule
\end{tabular*}
\begin{tabnote}
$^a$ \cite{Porco1995, Karkoschka2003}; $^b$ \cite{Dumas1999}; $^c$ \cite{dePater2005}. $^d$ Assumed: 15 km radial extent; 25$^\circ$ in azimuth; 
normal optical depth $\tau=0.1$ \citep{Porco1995}.
The NICMOS reflectivity is for the entire ring arc system.

\end{tabnote}
\label{tbl:tab3}
\end{table*}

Imaging, occultation and {\it in situ} data can be used to derive
particle properties of Neptune's rings.  The premier data set to 
determine such properties is the Voyager data set, because it is the
only one where both low- and high-phase angle data have been 
obtained. Particle properties as derived from the Voyager data have
been reviewed in detail by \cite{Porco1995}.
The particles in Neptune's rings are very dark, perhaps as dark as the particles
in the uranian rings. The fraction of optical
depth due to micrometer-sized dust is very high, $\sim$50\%, and
appears to vary from ring to ring. The limited data available are not
sufficient to make even an order-of-magnitude estimate of the mass of 
Neptune's rings, although the data suggest that the rings are
significantly less massive than the rings of
Uranus, unless they contain a substantial
population of undetected large ($\gtrsim 10$ m) particles, which is
unlikely given the paucity of smaller macroscopic ring particles.

Since the Voyager flyby, the data base has been extended by
groundbased near-infrared (\ref{sec2.1}) and visible light HST data
(\ref{sec2.2}).  Table~\ref{tbl:tab3} summarizes measured
reflectivities for Galatea and the ring arcs. As shown, Galatea has a
slight red color, whereas the ring arcs are very red. A red color is
typical for dusty rings, such as Jupiter's rings \citep{dePater1999},
Saturn's G ring \citep{dePater2004}, and the $\nu$ ring of Uranus
\citep{dePater2006}.

\section{Motion of the Arcs}\label{sec4}

\cite{Nicholson1995} analyzed the Voyager images together with $\sim$
25 stellar occultations by the Neptune system which were observed from
Earth-based telescopes between 1984 and 1988. All observations
combined indicated a mean orbital motion of the arcs of either
820.1194 $\pm$ 0.0006$^\circ$/day (referred to as solution 1) or 820.1118
$\pm$ 0.0006$^\circ$/day (referred to as solution 2). Solution 1 appeared
to be consistent with the value that was expected if the arcs are in a
42:43 corotation inclination resonance (CIR) with nearby satellite Galatea, a model
suggested soon after the arcs were first discovered in occultation
profiles \citep{Goldreich1986}, and attributed to the neptunian system
after the Voyager flybys by \cite{Porco1991}.

Galatea's average mean motion is $n_G = 839.66129 \pm 0.00002^\circ$/day. 
The mean motion of the 42:43 CIR with Galatea, $n_{CIR}$, creates 86 equally spaced
corotation sites around Neptune, which are located at potential
maxima. The CIR mean motion is given by
$n_{CIR} = (42 n_G + \dot{\Omega}_G)/43$, where $\dot{\Omega}_G$ is
Galatea's nodal precession rate. Using $\dot{\Omega}_G = - 0.714836^\circ$/day
\citep{owen91}, $n_{CIR} = 820.11760 \pm 0.00008^\circ$/day,
where the uncertainty is caused by that on Galatea's
mean motion \citep{Renner2014}.

Since Voyager, the ring arcs were imaged with HST/NICMOS
\citep{Dumas1999} and from the ground with CFHT \citep{Sicardy1999}, both in 1998. \cite{Dumas2002}
derived a mean motion of $820.1122 \pm 0.0003^\circ$/day for the arcs
from the HST data, and \cite{Sicardy1999} derived a value of $820.1135
\pm 0.0009^\circ$/day from the CFHT data. Using the Keck data taken in 2002 and 2003, 
\cite{dePater2005} derived a
value of $820.1118 \pm 0.0001^\circ$/day,
and \cite{Renner2014} determined a value of $820.11213 \pm 0.00008^\circ$/day
from VLT data taken in 2007. \cite{Showalter2013} obtained an almost identical
value, $820.1121 \pm 0.0001^\circ$/day, from HST/WFC3 observations. Clearly, the motion of the ring
arcs appears to be inconsistent with the value expected if the arcs are confined
by the above-mentioned 42:43 CIR.

The drift in mean
motion between the CIR and the arcs is
$\Delta n = n_{CIR} - n_{arcs} = (5.5 \pm 0.1) \times 10^{-3}$~$^\circ$/day,
equivalent to a mismatch $\Delta a \simeq 300 \pm 5$ m
in semi-major axis \citep{Renner2014}. This drift translates into a $\simeq 36 \pm 0.7^\circ$
difference for the arcs' longitude over the 18 years between the
Voyager and 2007 VLT data.  Given that all measured values for
$n_{arc}$ are essentially equal within their errorbars, it is clear
that the stability of the arcs cannot be explained with the CIR model.

In the following Section we will review the theory how to stabilize the arcs, and discuss 
the implications for Neptune's arcs.

\section{Stability of Arcs}\label{sec5}

\subsection{Resonant confinement}

\subsubsection{Spreading of a free Keplerian arc}

Stable ring-arcs are \textit{a priori} fragile structures, as
they must resist  both radial and azimuthal spreading.
Free arc particles moving along Keplerian orbits spread azimuthally at a rate 
\begin{equation}
\delta n = - \frac{3}{2} n \frac{\delta a}{a} \rm{,}
\end{equation}
where $\delta n$ is the difference in mean motion for a set of ring particles with a spread 
$\delta a$ in the semi-major axes, and where $n$ and $a$ are the mean orbital angular frequency
and the mean ring semi-major axis, respectively. The corresponding spreading timescale is
\begin{equation}
T_{S} = \frac{2 \pi}{|\delta n|} = \frac{4\pi}{3\sqrt{GM}} \frac{a^{5/2}}{\delta a} \rm{,}
\end{equation}

using Kepler's third law $a^3 n^2 = GM$. 

Thus, Neptune's ring arcs would be completely 
destroyed on a timescale $T_{S} \simeq 3.4$ years, using $a=62932.7$~km \citep{Renner2014}, 
$GM=6.8351  \times 10^6$ km$^3$~s$^{-2}$ \citep{owen91}, and assuming that the radial 
width of the arcs, $W \simeq 15$ km \citep{Hubbard1986b, Sicardy1991, Porco1995}, is equal to 
the semi-major axis dispersion $\delta a$.  
Actually, the Keplerian shear in a ring of width $W \simeq 15$ km and mean 
mean motion $n=820.11213^\circ$/day \citep{Renner2014} implies 
$\delta n \simeq - 0.29321^\circ$/day, i.e. a lengthening of the arcs 
of about $0.3^\circ$/day. Such a large value should have been detected by the
observations, thus showing that the arcs must be actively confined. 

All the arc models proposed so far essentially rely on the notion of 
corotation resonances (see Chapters 11 and~13).
An archetype of it is the co-orbital (or 1:1) mean motion resonance, 
in which the two triangular Lagrange points $L_4$ and $L_5$ offer
dynamically stable points  around which particles can librate.
The main problem with the Lagrange points, however, is that they correspond to local maxima of energy, 
and are therefore unstable against inter-particle, dissipative collisions.

Near a $m$+1:$m$ Mean Motion Resonance (MMR), there are actually two coupled resonances, 
one of corotation type (eccentric or inclined) and one of eccentric Lindblad type, 
see more details below, as well as in \cite{foryta96} and \cite{elmoutamid14}.
In that context, the same satellite can at the same time create 
the corotation sites where the arcs are azimuthally confined, and 
provide the energy necessary to counteract the effect of dissipative collisions,
thus confining the arcs radially.
This coupling is described in more detail in the next subsection.

\subsubsection{Corotation vs. Lindblad resonances}
\label{cor_lin}

In the restricted three-body problem (a massless particle perturbed by a satellite),
a corotation resonance
occurs at a radial location where the particle's mean motion $n$ equals the 
pattern speed $\Omega_p$ of a perturbing potential. 
Such a resonance excites periodic variations in semi-major axis $a$ and 
in longitude $\lambda$, leading to slow librations around equilibrium points.
In the frame rotating at angular velocity $n$, 
the particle guiding center has a pendulum-like motion along a closed loop that surrounds
one of the equilibrium points. 
This loop is enclosed into a separatrix, which defines a so-called corotation site.

In the example of the Lagrange points $L_4$ and $L_5$ mentioned earlier,
the pattern speed is simply the mean motion of the perturbing satellite, $n_S$.  
However, a satellite moving on an eccentric or inclined orbit creates additional 
corotation resonances located outside its own orbit. 
The strongest of them correspond to first order MMRs and 
have pattern speeds $\Omega_p$ that satisfy:  
\begin{equation}
 m \Omega_p = (m+1) n_S - \dot{\varpi}_S,
 \label{equ_C_loc}
\end{equation}
or 
\begin{equation}
 m \Omega_p = (m+1) n_S - \dot{\Omega}_S,
\end{equation}
where $\dot{\varpi}_S$ and $\dot{\Omega}_S$ 
are the apsidal and nodal satellite precession rates, respectively.
These two equations define, through $n=\Omega_p$,  
the Corotation Eccentric Resonance (CER) and the
Corotation Inclined Resonance (CIR), respectively.
For instance a CER corresponds to $\displaystyle n=\Omega_p=n_S+\kappa_S/m$, 
where $\kappa_S=n_S-\dot{\varpi}_S$ is the satellite epicyclic frequency.
Then, the mean motion of the particle matches the 
pattern speed $\Omega_p$ of the perturbing potential, hence the denomination 
corotation.
Finally, it can be shown that the CER creates $m$ 
corotation sites where arcs can potentially be maintained, while the CIR creates $2m$ sites \citep{Goldreich1986}.

To first order in $e$,
the particle motion can be described as the superposition
of an epicyclic motion of radial amplitude $ae$ and frequency $\kappa$
around a guiding center that moves uniformly at frequency $n$ 
along a circular orbit with radius $a$. 
The epicyclic frequency $\kappa$  and the mean motion $n$ are related by
$\kappa=n-\dot{\varpi}$, 
where $\dot{\varpi}$ is the particle apsidal precession rate.

The strongest (first order) Lindblad Eccentric Resonances (LERs) 
occur at radial locations where:
\begin{equation}
 m n = (m+1) n_S - \dot{\varpi} \rm{,}
 \label{equ_L_loc}
\end{equation}
or equivalently
\begin{equation}
 \kappa = (m+1) (n-n_S) \rm{.}
\end{equation}
Then, the epicyclic frequency is a multiple integer of the difference
between the particle's mean motion $n$ and the pattern speed, 
here $\Omega_p = n_S$, of the perturbing potential.
%

The quantity $n-n_S$ is the synodic frequency, i.e. the frequency at which 
the satellite is in conjunction with the particle and perturbs it through 
periodic tugs in the radial direction. 
The relation $\kappa = (m+1) (n-n_S)$ shows that at a LER,
the free epicyclic frequency of the particle matches one harmonic 
of the radial forcing due to the satellite.
Consequently, the LER excites the particle orbital eccentricity $e$.
This is in fact the source of energy that the particles need to balance the effect
of dissipative collisions. 

For $m \gg 1$, each corotation resonance is situated very near a
Lindblad resonance of the same satellite. 
Considering first the case where the particle is trapped in a CER and
perturbed by the associated LER, we have, from  
Eqs. (\ref{equ_C_loc}) and (\ref{equ_L_loc}):
\begin{equation}
 n_{CER} - n_{LER} = \frac{\dot{\varpi} - \dot{\varpi}_S}{m}.
\end{equation}
We use here the approximate expression $\dot{\varpi}\simeq 3n(R/a)^2 J_2/2$ for
the particle precession rate (with similar expression for the satellite),
where $R$ and $J_2$ are respectively the radius and the dynamical oblateness of the central planet.
Thus:
\begin{equation}
 a_{LER} - a_{CER} \simeq \frac{2}{3} \frac{a}{n} (n_{CER} - n_{LER})
 \simeq J_2 \Big{(} \frac{R}{a} \Big{)}^2 \frac{4a}{3m^2},
\end{equation} 
which shows that the CER is always \textit{inside} the LER, 
regardless of the position of the particle with respect to the satellite orbit
(i.e. regardless of the sign of $m$).

The same kind of exercise for a particle trapped in a CIR and perturbed by the
associated LER yields:
\begin{equation}
 a_{LER} - a_{CIR}  
 \simeq J_2 \Big{(} \frac{R}{a} \Big{)}^2 \frac{2a}{m}.
\end{equation} 
The situation is different from the previous case: 
the LER is now always \textit{between} the CIR and the satellite orbit.

The relative position of the corotation and Lindblad resonances
is an important ingredient for the arc confinement.
As discussed in Section \ref{nrg_transfer}, the energy transfer from the LER actually depends 
on the gradient of torque from the satellite, whose sign depends in turn 
on the position of the LER with respect to the corotation resonance.

We now review the different models for the arcs' confinement that have been developed 
to account for the available observations. 

\subsection{Arcs maintained by corotation resonances of a single satellite}

\cite{Goldreich1986} built up the most economical model for the arcs, requiring only one
satellite that perturbs the arcs through a $m$+1:$m$ first order MMR, 
as described in Section \ref{cor_lin}. 
%
%
%
A satellite on an inclined orbit rather than eccentric was originally proposed, 
because the eccentricity is more efficiently damped due to tidal effects. 
The perturbing potential arising from the corotation resonance is 

\begin{equation}
\Phi_{CIR}=-i_m^2 A_{CIR} \cos \Psi_{CIR}
\label{potcorot}
\end{equation}
where
\begin{equation}
A_{CIR}= {M_S \over M_P}n_S^2 a_S^2 V, 
\end{equation}
\vspace{-0.4cm}
\begin{equation}
V={\beta^2 \over 8} b_{3/2}^{(2 m+1)}(\beta),
\end{equation}
$i_m$  is the relative inclination between the satellite and ring orbits, 
$\Psi_{CIR}$ is the CIR resonance argument, $b_{3/2}^{(2 m+1)}$ is a Laplace
coefficient \citep{MurraySSD}, and $\beta =a/a_S$. \\

\cite{Porco1991} used the high resolution images from the Voyager spacecraft
to determine accurately the arcs' radial position, and explain their 
stability thanks to the resonant perturbations attributable 
to the nearby satellite Galatea. 
\cite{Porco1991} applied the model of \cite{Goldreich1986} and argued that the 
arcs are confined by a $42:43$ CIR with this moon\footnote{The resonance argument is 
$\Psi_{\rm CIR} = 2 [43 \lambda - 42 \lambda_G - \Omega_G]$, where $\lambda$ and 
$\Omega$ are respectively the mean longitude and the longitude of the ascending node, 
and the subscript $G$ denotes Galatea.}, using the mean motion of the 
arcs ($820.1185 \pm 0.0004^\circ$/day) derived by \cite{Nicholson1990}. 
More precisely, the location of the $42:43$ CIR was coincident with
the value of the semimajor axis ($a = 62932.37 \pm 0.02$) inferred from the arcs' mean
motion, considering the uncertainties and the CIR resonance
width (see Table~\ref{tbl:tab4}).
%
%
The Voyager data also indicated that the arcs have a radial distortion of amplitude 
$29.6 \pm 1.5$ km, compatible with the eccentricity forcing 
of the ring particles ($e_f=(4.7 \pm 0.2) \times 10^{-4}$) 
due to the associated $42:43$ LER, whose location 
falls 1.5 km interior to the ring's centerline (Table~\ref{tbl:tab4}). 
\cite{Porco1991} verified that the longitude at epoch of conjunction between
Galatea and the ring particles coincides with the longitude of the satellite, 
as expected if all of the particle orbits are exterior to the LER resonance. 
The mass of the satellite Galatea could be estimated from the observed forced eccentricity: 
$M_G = 2.12 \pm 0.21 \times 10^{18}$ kg.
Utilizing measurements of Galatea's size obtained from Voyager image analysis
\citep{Thomas91}, one obtains a density of $\rho_G=1.0 \pm 0.5$ g.cm$^{-3}$, 
a result falling within the range of reasonable densities expected in the Neptune
system.  

\begin{table}
\caption{42:43 Resonances$^a$ with Galatea.}
\begin{tabular}{ccc}
\toprule
Type &Location (km) & Width (km) \\
\hline
$42:43$ CIR &$62932.48 \pm 0.13$ & 0.6 \\
$42:43$ LER & $62930.83 \pm 0.13$ & 9 \\
\botrule
\end{tabular}

{$^a$ From \cite{Porco1991}.}
\label{tbl:tab4}
\end{table}

If the arcs are within the corotation sites of Galatea's 42:43 CIR, then one might expect them
to have azimuthal lengths less than $360/2|m|=4.1860^\circ$. The Voyager brightness profiles
showed that the arcs Courage, Libert\'e and Egalit\'e approximately meet this criterion, 
but not Fraternit\'e, whose angular extent was $9.6^\circ$. The arc separations are also
not exact integer multiples of $4.1860^\circ$. \cite{Smith1989} and \cite{Porco1991} suggested 
that bigger bodies, sources of the observed dust material, may librate in the corotation sites
and alter the position of the arcs. 

The CIR resonance width ($W_{CIR} \simeq 0.6$ km) is much
smaller than the observed width of the ring arcs (15 km). 
A low spread in semimajor axis $\Delta a$ is compatible with the coherent radial distortion  
observed through the ring. 
Otherwise, with e.g. $\Delta a \gtrsim 2$ km, the ring particles would
be located on either side of the Lindblad resonance, the orbits with $a>a_{\rm LER}$ and $a<a_{\rm LER}$
would be shifted by $\pi$, and therefore the radial excursion would not be coherent. 
On the other hand, if we assume that the spread in particle semimajor axes is equal to
the resonance width, $\Delta a = W_C$, then all particles would
have $a>a_{\rm LER}$, their perturbed orbits would all be in phase, and the
spread in their forced eccentricities, $\Delta e_f$, due to their differing
distances from the Lindblad resonance would yield a ring whose width would vary from 
$W_C$ at quadrature (where the particle orbits cross) to $a \Delta e_f$, 
which can be as large as the observed width of the arcs.
Then the arcs have more likely a radial width which results from the spread in eccentricity.\\



The \cite{Porco1991} model required revision after post-Voyager observations showed that the arcs' motion is not consistent with the value corresponding to the CIR
(Section \ref{sec4}), \cite{Namouni2002} showed how the 42:43 CER\footnote{The resonance argument is 
$\Psi_{CER} = 43 \lambda - 42 \lambda_G - \varpi_G$.}
can match the current arcs' semi-major axis and 
stabilize the system, if the arcs have a sufficient fraction of the mass of 
Galatea. This resonance creates 43 potential maxima, each of length $8.37^\circ$, 
which does not completely account for the angular lengths of the arcs. 
The ring mass determined to shift the CER to the arcs' position, $\simeq 0.002$ 
Galatea's mass, assumes an eccentricity of $10^{-6}$ for Galatea and would
correspond to a small satellite of 10 km in radius, for a density of 
$\simeq$ 1 g.cm$^{-3}$. The mass required in this model cannot be contained in a 
single body since Voyager data excluded undetected satellites of radius 
larger than 6 km (\cite{Smith1989}; \cite{Porco1995}). 
The exact origin of the small residual orbital eccentricity of Galatea,
consistent with a forcing by Adams ring's small mass, 
has to be determined too.

\subsection{Stability by co-orbital moonlets}

An alternative solution is to assume that the Adams ring is a collection of a few 
moonlets that  maintain stable co-orbital relative positions akin to the Lagrangian 
$L_4$/$L_5$ points \citep{renner04}.
The observed arcs would then be composed of dust trapped between 
those co-orbital satellites, an intermediate situation between a
fully collisional ring with only small particles, and a fully accreted system
of only one satellite. 
This generalizes the original shepherding model proposed by \cite{lissauer85}
and is an adaptation of \cite{sicardy92}. 

Indeed, \cite{lissauer85} proposed that two satellites could stabilize arcs, 
one maintaining particles at the $L_4$ and/or $L_5$ points, and another one 
providing the energy lost through collisions, confining radially the arc material thanks
to the Lindblad resonances described in Section \ref{cor_lin}.
This model predicts only one stable arc associated with the Lagrangian satellite. 
Despite the discovery of the satellite Galatea located $\sim 980$ km interior to the arcs,  
the images of the Voyager 2 mission, for which the maximal resolution was 
12 km \citep{Smith1989}, did not reveal any co-orbital moon in the Adams ring.

\cite{sicardy92} proposed a variant model by considering two co-orbital 
satellites at their mutual $L_4$ point, which allows the possible existence of 
satellites smaller than in the \cite{lissauer85} model. 
%
%
However, the detailed geometry of Neptune's ring arcs (four arcs with different
angular lengths) is still unexplained with this model, requiring more sophisticated
solutions.

\subsubsection{Family of stable stationary configurations}
  
Results on the existence of stationary configurations for co-orbital
satellites with small and arbitrary masses were derived by \cite{renner04}. 

Consider the circular planar restricted $1+N$ body problem with $N$ co-orbital 
satellites moving with the same average mean motion $n_0$ (and orbital
radius $r_0$) around a point mass $M$. In a coordinate system with origin at
$M$ and rotating with the velocity $n_0$, the position of a satellite 
is described by the coordinates $(\phi_i,\xi_i)$,
$i=1,...,N$, where $\phi_i$ is the longitude of satellite $i$ with
respect to an arbitrary origin, and $\xi_i=\Delta r_i /r_0$ is the
relative radial excursion of that satellite. Choosing units such that $n_0 = r_0 = M=1$, 
the equations of motion are then (see also \cite{salo88}) 

\begin{equation}
\left\{
\begin{array}{lr}
\displaystyle  {\dot \phi_i}= -{3 \over 2} \xi_i \\ \\ 
\displaystyle {\dot \xi_i}= -2  \sum_{j \neq i} m_j f'(\phi_i - \phi_j)
\end{array}
\right.
\label{ncoorb}
\end{equation}
where
\begin{equation}
 f(\phi)= \cos \phi - {1 \over 2|\sin \phi/2|} ,
 \label{deff}
\end{equation}

In order to ensure that the satellites remain co-orbital, 
this system also assumes that any two satellites $i$ and $j$ never get closer 
than a few mutual Hill radii
$r_H = r_0 \cdot \left[\left(m_i + m_j\right)/M\right]^{1/3}$. 
The first equation in (\ref{ncoorb}) represents the differential
keplerian velocity of each satellite with respect to $r_0$, 
and the function $f(\phi)$ in Eq.~(\ref{deff}) is the sum of the indirect and
direct potentials exerted by a given satellite on the other co-orbital
bodies. 
The fixed points of system (\ref{ncoorb}) are given by:

\begin{equation}
\xi_i=0
\label{equi1}
\end{equation}
and
\begin{equation}
\sum_{j \neq i} m_j f'(\phi_i - \phi_j)=0
\label{equi2}
\end{equation}

for all $i=1,...,N$. 
Eq.~(\ref{equi1}) means that in a stationary configuration the $N$
co-orbital satellites have the same orbital
radius. 
Eq.~(\ref{equi2}) involves the angular separations between the
satellites and can be written in a matrix form. Defining $f^\prime_{ij}
\equiv f'(\phi_i-\phi_j)$ and noting that the function $f^\prime$ is odd, we
obtain from Eq.~(\ref{equi2}):

\begin{equation}
\left [ \begin{array}{ccccc} 
0 & f'_{12} & \cdots & \cdots & f'_{1N} \\ 
- f'_{12} & 0 & f'_{23} & \cdots & f'_{2N} \\ 
\vdots &   & 0 &   &  \\
\vdots &  &  & \ddots &  \\
- f'_{1N} &  &  &  & 0
\end{array} \right]
\cdot
\left [ \begin{array}{ccccc} 
m_1 \\ m_2 \\ \vdots \\ \vdots \\ m_N
\end{array} \right]  =
0_{{\mathbb R}^N} 
\label{solnpar}
\end{equation}

Since the matrix $M$ above is antisymmetric, its rank is even. 
Consequently, for given arbitrary angles $\phi_i$'s,
the existence of non-trivial solutions (positive or not) of the linear system 
(Eq.~\ref{solnpar}) depends on the parity of the number $N$ of satellites.
If $N$ is odd, then there always exists a set of mass values 
that achieves stationarity for any arbitrary angular separation between the satellites.
However, strictly positive masses restrict this existence to sub-domains of angular separations. 
If $N$ is even, the rank of $M$ is generally $N$, 
and there is in general no family of mass vectors that leads to a stationary configuration. 
This is well known for the case $N=2$, where only the triangular
points $L_4$ and $L_5$ and the diametral point $L_3$ allow the
satellites to be stationary with respect to each other. Thus, one must
first cancel the determinant of the matrix, det$(M)$, in order to
obtain non-trivial solutions of (Eq.~\ref{solnpar}).
%

\begin{figure}
\figurebox{}{6pc}{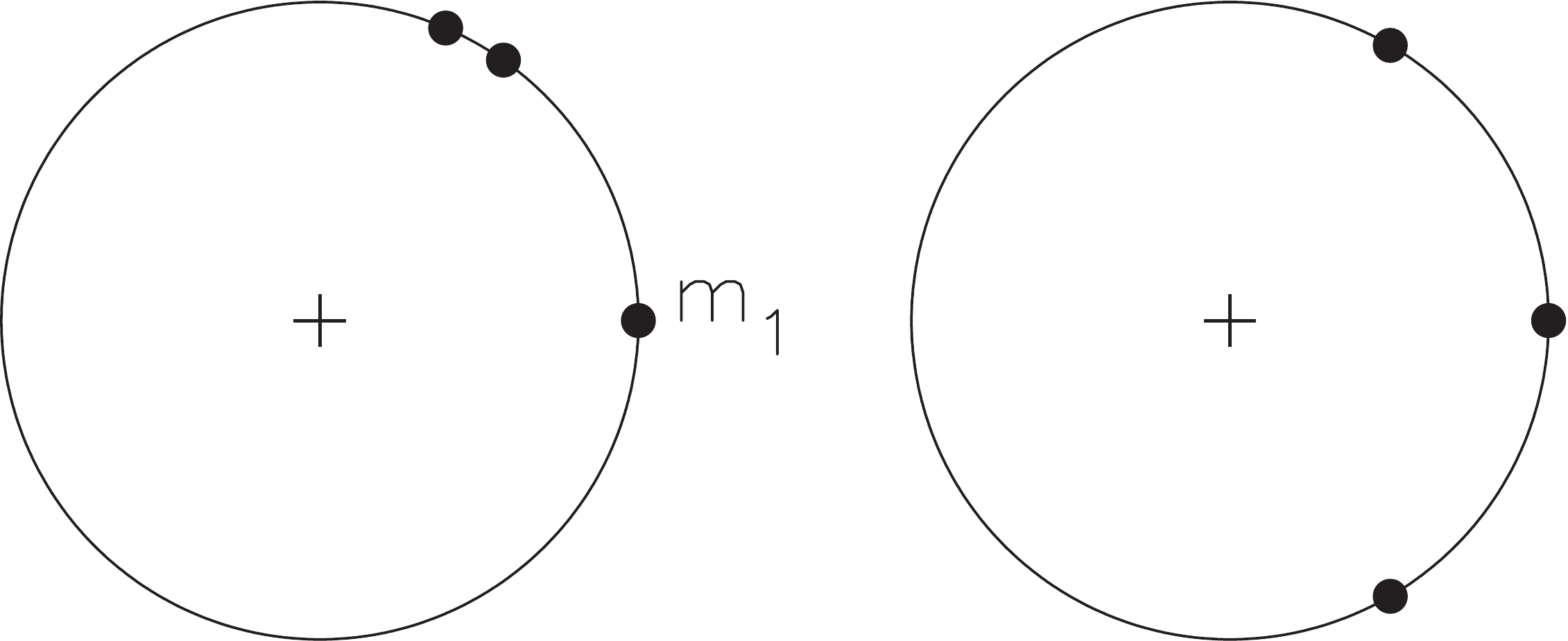}
\caption{Stable stationary configurations for $N=3$ co-orbital satellites with
$m_2=m_3=10^{-2}m_1$. 
From left
to right, the angular separations of satellites 2 and 3 with respect
to satellite 1 are: 
$(\phi_2,\phi_3)= (54^\circ.84,66^\circ.75)$,
$(59^\circ.82,300^\circ.18)$. (Figure from \cite{renner04})}
\label{3N_1}
\end{figure}

\cite{renner04} showed that the case $N = 3$ can be completely treated analytically for 
small arbitrary satellite masses, 
giving all the possible solutions and their linear stability. For $N \geq 4$, 
the possible stable stationary configurations for given satellite masses are derived 
following a numerical approach where the satellites converge towards the linearly stable equilibria, 
by adding a non-conservative term which brings energy in the equations of motion. 
Integrating these perturbed equations and exploring
random initial coordinates with random masses then provides the domains
of stable stationary points. These domains 
correspond to configurations where the co-orbitals are either positioned near 
the $L_4$ and $L_5$ points of the most massive satellite, or are grouped near one of 
these two points, as shown in the example displayed in Fig.~\ref{3N_1}.

\subsubsection{Application to observations}

\begin{figure}
\figurebox{}{16pc}{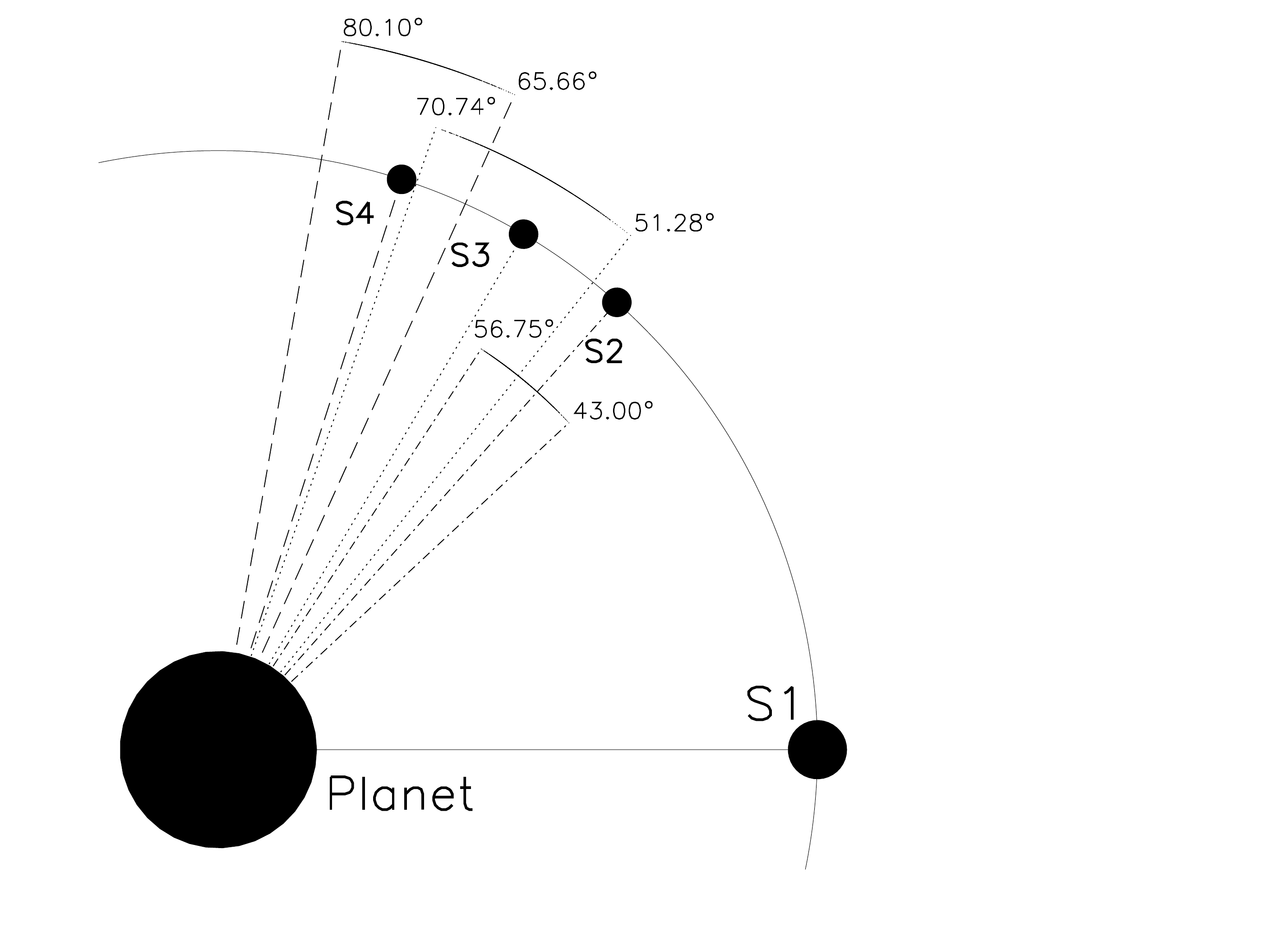}
\caption{Subset of stationary configurations for $N=4$ satellites, akin 
	  to the Lagrangian $L_4/L_5$ points.  
	  The mass ratios $m_i/m_1$ of the satellites 
	  $S_i$ ($i=2,3,4$) with respect to the most massive $S_1$ are 
	  lower than 0.03, and the angular separations 
	  are compatible with the observed inter-arc regions
	  (see text). The satellites in equilibrium are
	  close to the $L_4/L_5$ point of $S_1$:$S_2$ at a longitude with respect 
	  to $S_1$ between $43$ and $56.75^\circ$, $S_3$ between $51.28$ and $70.74^\circ$ 
	  and $S_4$ between $65.66$ and $80.1^\circ$. (Figure from \cite{Renner2014})}
	 \label{coorb_family}
\end{figure}

\cite{Renner2014} demonstrated how the 
observed inter-arc regions lead to a limited space of possible mass ratios between 
hypothetical co-orbital satellites, achieving equilibrium.  
They applied the method above to $N=4$ satellites (where the satellite $i=1$ is 
by convention the most massive) with random initial mass ratios
$m_i/m_1$ ($i=2,3,4$) lower than three percent\footnote{Increasing this value 
increases the average relative angular positions of the solutions, 
which could lead to configurations incompatible with the observed inter-arc regions.}, 
and stored the stable 
solutions with angular separations satisfying $\Phi_{32}=11 \pm 3^\circ$, 
$\Phi_{43}=12.5 \pm 3^\circ$ and $\Phi_{42}=23.5 \pm 3^\circ$ (where 
$\Phi_{ij}$ is the azimuthal separation between the satellites $i$ and $j$), 
that is, configurations compatible with the observed inter-arc regions
(Porco, 1991; Dumas et al., 2002; de Pater et al., 2005). 
A continuum of mass values is obtained, with angular configurations where the 
co-orbitals are near the $L_4$ (or $L_5$) point of the most massive satellite, 
as illustrated in Fig. \ref{coorb_family}.
%
Full numerical integrations then confirmed that the system is numerically 
stable under the effects of the perturbations due to the satellite Galatea.
These perturbations are mainly the 42:43 LER and the 42:43 CIR, located about 0.25 km
inside the arcs' semimajor axis. The moonlets have a slow libration motion around the 
stationary configuration, and arc particles are strongly maintained between the 
co-orbital satellites on much longer time scales than the observation time span 
of $\sim 30$ years. 

\subsection{Energy and angular momentum transfer}
\label{nrg_transfer}

As mentioned earlier, 
the $m$ (resp. $2m$) stable equilibria distributed around the orbit of a
corotation resonance are at potential maxima. Thus, the energy dissipated by interparticle
collisions must be replenished for an arc to remain confined. 
In general, a LER tends to exert a repulsive torque $\Gamma$ on the arc particle,
i.e. it tends to pull apart the particles and the satellites, independently
of the detailed physics at play in the ring\footnote{
This \textit{standard} torque can be written 
\begin{equation}
\Gamma = 0.4 a^2 n^2 \Big{(} {M_S \over M_P} \Big{)}^2 \Big{(} {a \over a-a_S} 
\Big{)}^4 M_r, 
\end{equation}
where $M_r$ is the mass of the ring.} \citep{goldreich80, meyer87}.
Thus, 
arcs orbiting outside a satellite ($m<0$) should gain angular momentum (and thus orbital energy), while 
arcs orbiting inside ($m>0$) should lose orbital energy. 
Naively, one may thus expect that only arcs orbiting outside a satellite can be maintained.
Things are more subtle, though, because a \textit{differential} torque must in fact 
be exerted between the inner and the outer edges of the arc for stabilizing it.
If we place ourselves in the frame corotating with the arcs, and if we
denote $\zeta$ the potential energy (per unit mass) associated with the corotation resonance,
then change rate $\dot{\zeta}$ for particles 
around corotation points, averaged over one libration period $T_C$, is
\begin{equation}
\dot{\zeta} = - \frac{3 n_S}{4 \pi \sigma a_0^2} 
\frac{1}{T_C} \int_{T_C} a \frac{d \Gamma}{da} dt,
\label{energy_transfer}
\end{equation}
where $a_0$ is the average orbital radius, $\sigma$ the ring surface density, 
and $\displaystyle d \Gamma / da$ the torque density, that is, the torque exerted 
by the satellite per unit radius \citep{Sicardy1991}. 
Because of the presence of the term $a$ in front of $\displaystyle 
d \Gamma / da$, the energy received is proportional
to the gradient $\displaystyle d^2 \Gamma / da^2$ of the torque density. 
Physically, this means that a given particle must undergo a larger repulsion
on the inner side of its libration path than on the outer side, in order 
to compensate for the spreading effect due to dissipation.
Thus, the resulting differential torque across a corotation site can serve to
stabilize particles and form a stable ring arc. 

\cite{sicardy92} noticed that most of the energy supplied to 
the arcs is used to excite the eccentricities, not to confine the particles, 
assuming that the standard torque formula holds.
%
%
%
Assuming that a stationary state is reached, the orbital migration of the Adams ring 
is approximately 3 km/yr, a value too high to stabilize the ring 
on long time scales. 
However, the standard torque formula may not be applicable for Neptune's arcs since it assumes 
that Lindblad resonances $m+1:m$ overlap,  whereas the radial separation between two resonances is 
$\sim 25$ km, i.e. larger than the physical width of the ring.

\subsection{Collisions}

The most critical problem for the arc stability is the inelastic collisions 
between particles. 
In the framework of the CIR model with Galatea, \cite{foryta96} noticed 
that the forced eccentricity gradient due to the Lindblad resonance 
implies potential collisions with typical relative velocities of about 
1 m.s$^{-1}$, which can cause semimajor axis changes of about 10 km. 
Impacts between macroscopic particles, necessary to reproduce the observed dust 
ratios, break the resonance capture with the satellite over short time scales,
as confirmed by numerical collisional N-body simulations \citep{hanninen97}.
%
%
%
\cite{salo98} then considered self-gravity between macroscopic co-orbital particles
(with radii in the range 0.1-1 km) to effectively reduce the collision frequency
and thus stabilize the arcs in CIR with Galatea. 
The equations (\ref{ncoorb}), coupled with the CIR equations, were used 
to find that a few kilometer-sized particles
can be maintained in resonance without colliding with each other.
In this model, the kilometer-sized bodies separate the arc into different segments, 
which explains partly the observed azimuthal lengths, while sub-kilometer particles 
supply dust in the ring and account for the clumpy structure of the arcs.  
%

\section{Summary, Conclusions, and Outlook}\label{sec6}

Although much progress has been made on characterizing and
understanding Neptune's ring system, including its ring arcs, our
knowledge is clearly hampered by the scarcity of good data, which by
itself is not surprising as the rings and ring arcs are very faint.
Since the Voyager flybys, several data sets have been obtained with HST, Keck and the VLT; 
some of these data have been shown in this review chapter, but clearly
much work remains to be done. The data reduction and analysis,
including careful photometry, of these data is tedious and time
consuming. In the future, with the next generation of large, 30--40 m,
ground-based telescopes and JWST, the sensitivity to Neptune's rings
and ring arcs will be much improved. But rather than waiting for these
telescopes, it remains important during the
intervening years to continue observing the ongoing evolution of the ring
arcs; perhaps new arcs may form. The physics of confining these
arcs, and the original formation of the arcs can only be determined if
a database of the evolution of the ring arcs is established.

In addition to the ring arcs, it is also crucial to regularly observe
Neptune's small moons, as the interaction between moons and ring arcs
is clearly important to understand the physics. Based upon
observations taken so far, it is clear that the orbits of Neptune's
moons are not well-characterized by the Voyager data, and one new inner moon
has been discovered since then \citep{showaltercbet3586}. 

\subsection{Origin}

The origin of the arc system could be the breakup 
of a parent satellite or the accretion of ring material within the 
Roche zone of the planet. 
The latter is outside the Adams ring for a density of 1 g cm$^{-3}$
\citep{esposito02}, allowing a mixture of collisional rings and accreted moonlets.  
The Galatea secular torque (Section \ref{nrg_transfer}) could help to 
set a few ring moonlets in stable stationary 
configurations, provided that the energy generated by the Lindblad 
resonance is higher than that dissipated through collisions. 
The exact origin of the proposed configuration of co-orbital bodies is still unknown. 
Such a mechanism could be hierarchical if accretion plays a significant role:
a (previously accreted) satellite could gather some ring material around its 
$L_4$ or $L_5$ Lagrangian point, then larger co-orbital particles would form close
to $L_4/L_5$ until a stationary configuration is reached. 
The ring arcs
observed would be the residual material confined in between these small
satellites. Simulations of self-gravitating and colliding particles 
(see, e.g., \cite{rein12}) need to be performed to study the formation mechanisms 
of small satellites close to the Roche zone of a planet.  
Once the system of co-orbital bodies is in a stationary configuration, the
secular torque and therefore the ring orbital migration are reduced
\citep{sicardy92}.

\subsection{Dynamical evolution}

The dynamical evolution of Neptune's partial rings is unknown: 
the existence of a system of small co-orbital satellites might be 
part of an incomplete process of satellite formation. 
Alternatively, Neptune's arcs could be 
transitory, resulting in a new ring if the equilibrium configuration 
breaks down.
The latest Keck and HST observations show that the leading two arcs 
(Courage, Libert\'e) appear to have vanished, 
while the trailing two (Egalit\'e, Fraternit\'e) appear to have remained 
quite stable. Though it seems they must be submitted to confinement mechanisms, the arcs nevertheless evolve rapidly.

\section*{Acknowledgements}

I.d.P. acknowledges partial support from NSF grant 1615004.  
M.R.S. acknowledges support from NASA's Outer Planets Research Program through grant NNX14AO40G, 
and from Space Telescope Science Institute through program HST-GO-14217. 
Support for Program number HST-GO-14217 was provided by NASA through a grant from the Space Telescope Science Institute, 
which is operated by the Association of Universities for Research in Astronomy, Incorporated, under NASA contract NAS5-26555.  
B.S. acknowledges funding from the French grant "Beyond Neptune II" (ANR-11-IS56-0002) and 
from the European Research Council under the European Community's H2020 (2014-2020/ ERC Grant Agreement no. 669416 "LUCKY STAR").


\end{document}